\documentclass[aps,prl,reprint,groupedaddress]{revtex4-1}
\usepackage{braket}
\usepackage{graphicx}
\usepackage{hyperref}

\begin{document}

\preprint{}

\title{
Ultraflat bands and shear solitons in Moir\'e patterns of twisted bilayer transition metal
dichalcogenides.
}


\author{Mit H. Naik}
\author{Manish Jain}
\email{mjain@iisc.ac.in}
\affiliation{Center for Condensed Matter Theory, Department of Physics, Indian Institute of Science, Bangalore 560012, India}


\date{\today}

\begin{abstract}
Ultraflat bands in twisted bilayers of two-dimensional materials have
potential to host strong correlations, including the Mott-insulating phase at half-filling of the band.
Using first principles density functional theory calculations,
we show the emergence of ultraflat bands at the valence band edge in twisted bilayer 
MoS$_2$, a prototypical transition metal dichalcogenide.
The computed band widths, 5 meV 
and 23 meV for 56.5$^\circ$ and 3.5$^\circ$ twist angles
respectively, are
comparable to that of twisted bilayer graphene near 'magic' angles. 
Large structural transformations in the 
Moir\'e patterns lead to formation of shear solitons at stacking boundaries
and strongly influence the electronic structure.
We extend our analysis for twisted bilayer MoS$_2$ to show that flat bands can occur
at the valence band edge of twisted bilayer WS$_2$, MoSe$_2$ and WSe$_2$ as well.

\end{abstract}


\maketitle


Combining bilayers of two-dimensional materials with a small-angle twist between the layers 
or combining two dissimilar 2D materials with a small lattice mismatch leads to the formation of  
Moir\'e superlattices (MSL) with 
periodicity in the order of nanometers \cite{PNAS.Alden,NPhys.Woods,NPhys.Matthew} . 
MSL in twisted bilayer graphene (tBLG) host a plethora of 
fascinating physics at the structural \cite{PNAS.Alden,NMat.Matthew,NL.Jose} and electronic level \cite{NPhys.Li,PRB.Wong,PRB.Jose,NN.Koren,NL.Patel,PRL.Lado}.
Rearrangement of atoms in the MSL leads to the formation of
shear solitons and topological point defects \cite{2D.Oleg,PNAS.Alden,2D.Wijk,NL.Jose}. 
The electronic structure of these MSLs can be
different from that of the constituent layers, like formation of flat bands and localization of states 
close to the Fermi level \cite{2D.Oleg,NL.Kang,PRB.Kang,Nature.Cao,Nature.Cao2}. 
Probing flat bands in tBLG has recently led to the discovery of 
unconventional superconductivity close to a 'magic' angle.
\cite{Nature.Cao,Nature.Cao2,PNAS.MacD,PRB.Suarez}.

\begin{figure}
  \includegraphics[scale=0.245]{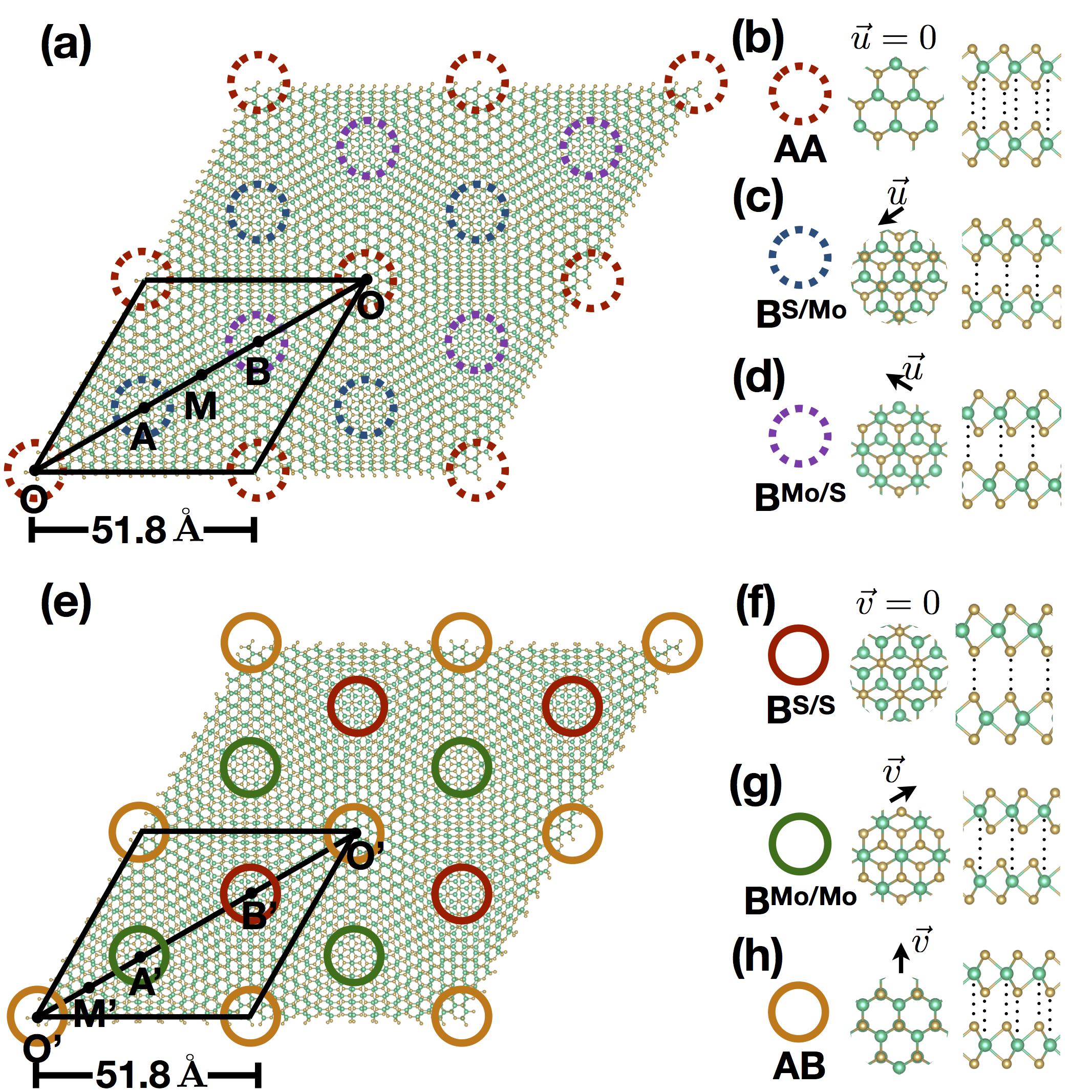}

 \caption{\label{fig1}
 (a) and (e) Moir\'e superlattice formed by twisting bilayer MoS$_2$ by 3.5$^\circ$ 
 and 56.5$^\circ$ respectively. The high-symmetry stackings are highlighted in these superlattices using 
 circles. The stacking within the circle is shown in (b)--(h). The direction 
 of the order parameter, $\vec{u}$ or $\vec{v}$ (see text for details), is also shown.
}
\end{figure}

MoS$_2$, a 2D transition metal dichalcogenide (TMD), is arguably the most 
popular 2D material after graphene \cite{PRL_Mak,NN_Radisavljevic,NN.Wang}. 
Due to its semiconducting nature
extensive applications in electronics and optoelectronics 
have been explored \cite{NN_Radisavljevic,NN_Roy,NN.Wang}. 
However, 
in contrast to tBLG, MSL
in twisted bilayer MoS$_2$ (tBLM) have not received as much attention \cite{NL.Zande,NL.Huang,NL.Huang2,NL.Yeh,NComm.Liu,Carr.twistronics,AN.Lin}.  
In this letter, we use first-principles density functional theory (DFT) \cite{PR.Kohn} calculations 
to study the electronic and structural transformations in the MSLs of tBLM. 
We show a large structural reconstruction
in the Moir\'e pattern, leading to the 
formation of shear solitons and ultraflat bands at the valence band edge of tBLM.
These flat bands have band widths comparable to those observed in tBLG close to 'magic' 
angles \cite{Nature.Cao,Nature.Cao2}.
Our calculations show that in-plane 
relaxations of the layers drive the out-of-plane relaxations, which 
in turn lead to localization of the VBM.
We show that the spatial localization of the flat band
significantly changes if the bilayers are rigidly twisted, and relaxations ignored.  
Localization of the flat band can influence exciton dynamics and binding energy in the 
Moir\'e pattern.  
The ratio of on-site Coulomb interaction to band width of the flat band is found to be 
large, indicating the 
possibility of a Mott-insulating phase at half-filling of the band.

tBLM forms two distinguishable MSLs for small twist angles close to 0$^\circ$ 
and close to 60$^\circ$. For tBLG, these are equivalent.
Fig. \ref{fig1} (a) and (e) show the MSLs formed for twist angle 
3.5$^\circ$ (M$^{3.5}$) and 56.5$^\circ$ (M$^{56.5}$), repectively. 
These superlattices are composed of 
various high-symmetry stackings. 
We define B$^\mathrm{X/Y}$ as being a bernal-like stacking of the two layers with 
X atom in the top layer directly above the Y atom in the bottom layer.
M$^{3.5}$ consists of the 
AA stacking, B$^{\mathrm{S/Mo}}$ and B$^{\mathrm{Mo/S}}$
(Fig. \ref{fig1}). 
M$^{56.5}$ similarly consists of the
B$^\mathrm{S/S}$, B$^\mathrm{Mo/Mo}$ and AB (Fig. \ref{fig1}) stackings. 
We note that no simple translation transforms the AA stacking to AB stacking.
We define an order-parameter, $\vec{u}$ \cite{2D.Oleg,PNAS.Alden}, for twist angles close to 0$^\circ$
as the shortest displacement vector that takes any given stacking to the highest energy
stacking in the corresponding Moir\'e pattern; AA stacking in this case.
For twist angle close to 60$^\circ$, we define $\vec{v}$, as the shortest 
displacement vector that takes any given stacking to the highest energy
stacking, B$^\mathrm{S/S}$. 

All the DFT calculations are performed using the 
pseudopotential plane-wave package, Quantum Espresso \cite{QE.Giannozi}.  
We simulate the following angles in this study: 3.5$^\circ$, 5.1$^\circ$, 7.3$^\circ$, 
56.5$^\circ$,54.9$^\circ$ and 52.7$^\circ$ \cite{PRB.Mele,PRL.Lopes}. 
The code Twister \cite{twister} is used to generate the atom positions for these structures. 
M$^{3.5}$ and M$^{56.5}$ are 
the largest systems in our calculation, containing 1626 atoms. 
In all MSL calculations, the Brillouin zone is sampled at the $\Gamma$ point to obtain the 
self-consistent charge density. The Hamiltonian is subsequently constructed and diagonalized 
at other k-points in the Brillouin zone to obtain the band structure.
(see Supplementary Materials (SM) for more details \cite{SM_ref}).

\begin{figure}
  \includegraphics[scale=0.24]{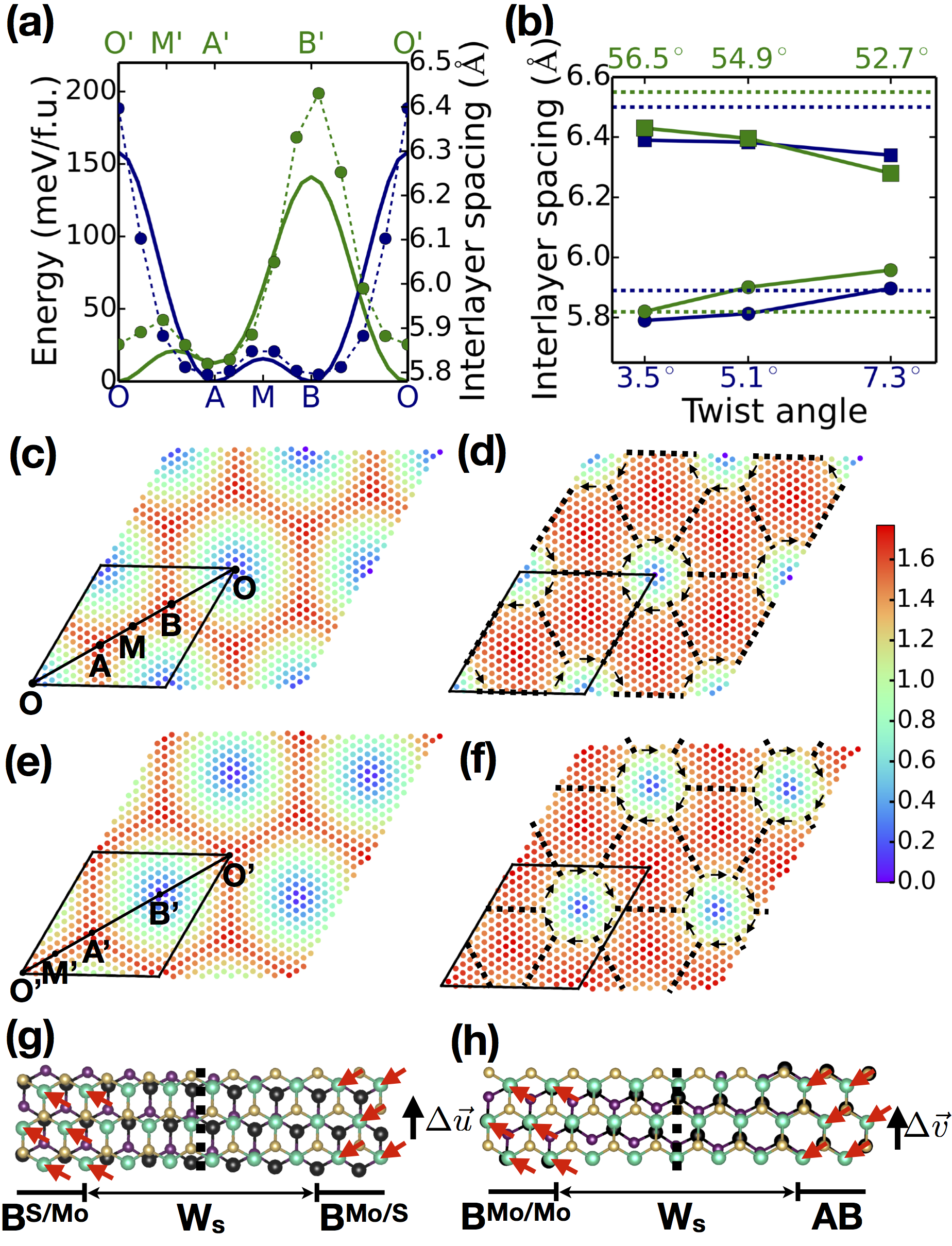}

 \caption{\label{fig2}
 (a) The relative energy and ILS of the stackings along a line in the MSL.
  The blue (green) lines corresponds to the path in M$^{3.5}$ (M$^{56.5}$). 
  The solid (dashed) lines represent the relative total energy (ILS) along the path.
 (b) Maximum (squares) and minimum (circles) ILS in the MSL as a function of twist angle.
     The green and blue lines correspond to angles approaching 60$^\circ$ and 
     0$^\circ$, respectively. The dashed lines represent the 
     maximum and minimum equilibrium ILS of the corresponding isolated stackings. 
  (c) and (e) ((d) and (f)) Distribution of $|\vec{u}|$ and $|\vec{v}|$ in the unrelaxed (relaxed) MSLs.
  The dashed lines denote stacking boundaries.
  The arrows in (d) and (f) denote the direction of order parameters.
  (g) and (h) Show the structure, order parameter (red arrows) and the change in order parameter 
  across a stacking boundary.
}
\end{figure}

To understand the relaxation in the MSL, we study the relative energies of the 
stackings
keeping the interlayer spacing (ILS) fixed at 5.9 $\mathrm{\AA}$.
The relative total energy per unit MoS$_2$ along the line traversing high-symmetry stackings
(defined in Fig. \ref{fig1} (a) and (e)) 
in M$^{3.5}$ and M$^{56.5}$ is shown in
Fig. \ref{fig2} (a). 
The AA (O point) and B$^\mathrm{S/S}$ (B' point) stackings have S atoms 
of the top layer directly above the 
S atoms in the bottom layer. Strong repulsion between the out-of-plane S-$p_z$ orbitals 
causes these stackings to have the highest energy.
Stackings with S on top of Mo are energetically favourable.
On relaxing the MSLs, large structural 
transformations with respect to the rigidly-twisted bilayer are observed. 
The out-of-plane displacements of these relaxation patterns
lead to local variations in the ILS between 
the two layers. 
Fig. \ref{fig2} (a) shows the ILS in M$^{3.5}$ and M$^{56.5}$ along the line 
defined in Fig. \ref{fig1} (a) and (e). The relative energies correlate 
strongly with the ILS. 
The variation in the ILSs increase as the twist angles approach 0$^\circ$ or 
60$^\circ$ (Fig. \ref{fig2} (b)).
The local ILS is maximum 
in the AA (B$^\mathrm{S/S}$) stacked patches for M$^{3.5}$ (M$^{56.5}$), 
and is comparable to that of the isolated AA (B$^\mathrm{S/S}$) stacking.
The ILS is smallest for the B$^\mathrm{Mo/S}$ (B$^\mathrm{Mo/Mo}$) and B$^\mathrm{S/Mo}$ stackings (AB) 
in M$^{3.5}$ (M$^{56.5}$).
This variation in ILS in tBLM ($\sim$0.6 \AA) is much larger than  
tBLG ($\sim$0.2 \AA) \cite{2D.Oleg}.
To understand the coupling between the in-plane and out-of-plane displacements, 
we relax the tBLM keeping the in-plane positions of the atoms fixed (ie. no in-plane shear) 
to that of the rigidly twisted bilayer.
On adding this constraint, the 
variation in the ILS across the MSLs is significantly reduced (see SM). 

The order parameter vectors, $\vec{u}$ and $\vec{v}$, are computed locally for 
every Mo atom in
 M$^{3.5}$ and M$^{56.5}$, respectively. The spatial variation of $|\vec{u}|$ and $|\vec{v}|$
for rigidly tBLM is shown in Fig. \ref{fig2} (c) and (e). $|\vec{u}|$ = 0 and $|\vec{v}|$ = 0 for O (AA) and 
B' (B$^\mathrm{S/S}$) regions, as defined. $|\vec{u}|$ takes a maximal value of 1.8 $\mathrm{\AA}$, 
ie. the in-plane Mo-S distance, for 
A (B$^\mathrm{S/Mo}$) and B (B$^\mathrm{Mo/S}$) regions. Similarly, $|\vec{v}|$ = 1.8 $\mathrm{\AA}$ for
O' (AB) and A' (B$^\mathrm{Mo/Mo}$) region. Thus large values of $|\vec{u}|$ and $|\vec{v}|$ indicate 
regions of low energy (Fig. \ref{fig2} (a))
On relaxing the structure, the MoS$_2$ units in the top layer and bottom layer displace in opposite 
directions in-plane forming a vortex-like pattern around the O and B' points (see SM). 
These in-plane displacements in M$^{3.5}$ and M$^{56.5}$ increase the fractional area of 
the respective low energy stackings. 
The large variation in the distribution of $|\vec{u}|$ and $|\vec{v}|$ as shown in Fig. \ref{fig2} (d) and (f) 
from the rigidly tBLM is indicative of this transformation.  

A domain boundary separates adjacent low energy stackings, denoted by the dashed lines in Fig. \ref{fig2} 
(d) and (f). 
A similar transformation is observed in tBLG for small twist angles, 
leading to the formation of triangular domains of Bernal stackings, with the AA stacking at the 
vertices of the triangles \cite{2D.Oleg,PNAS.Alden}.
A shift (change in the order parameter) in the relative registry of the atoms is necessary to traverse accross the 
boundary. 
This change in the order parameter is parallel to the domain boundary, 
indicating a shear-strain soliton \cite{PNAS.Alden}.
Fig. \ref{fig2} (g) and (h) shows the shear-strain 
soliton between B$^\mathrm{S/Mo}$ and B$^\mathrm{Mo/S}$
stackings in M$^{3.5}$, and between AB and B$^\mathrm{Mo/Mo}$ in M$^{56.5}$, respectively.
The width of the soliton is $\sim$1.5 nm, W$_\mathrm{s}$. The soliton width is 
expected to be larger for smaller angles and to ultimately saturate \cite{2D.Oleg} for vanishing angles.
Furthermore, the order parameter in the vicinity of the O ($\vec{u}$ = 0) and B' ($\vec{v}$ = 0) 
points rotates by 2$\pi$ (see Fig. \ref{fig2} (d), (f)). These points are thus 
topological point defects \cite{PNAS.Alden}.  

The electronic structure of the MSLs can be understood in terms of
 the constituent high-symmetry stackings. The band structure 
of BLM is sensitive to stacking and ILS. 
Fig. 3 shows the
band structure of the five high-symmetry stackings.
The VBM in all BLM stackings is at the $\Gamma$ point, unlike BLG where it is at the $K$ point.
The band structure close to the Fermi level, for all stackings, show 
large band splittings at the $\Gamma$ point and relatively 
small splittings at the K point. This is indicative of the strength of 
hybridization between the two layers at these points. 
The K and $\Gamma$ point wavefunctions close to the Fermi level
in monolayer 
MoS$_2$  have small and large spreads in 
the out-of-plane direction, respectively \cite{PRB.Naik}. 
This spread determines the hybridization, leading to
the different splittings.

\begin{figure}
  \includegraphics[scale=0.24]{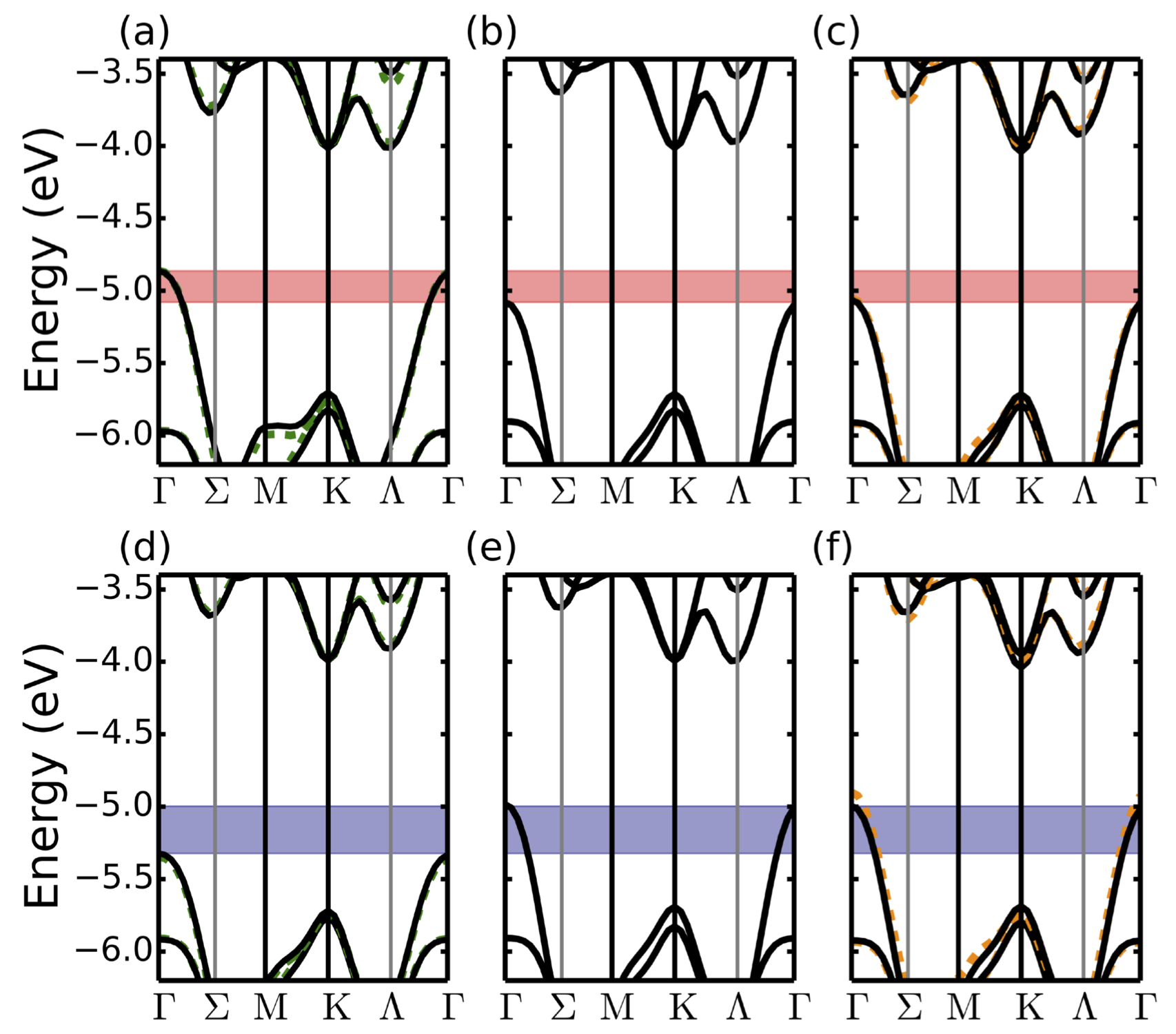}

 \caption{
 \label{fig3}
 (a), (b) and (c) Band structures of the isolated high-symmetry stackings with interlayer spacing 
 fixed at 5.9$\mathrm{\AA}$. AA and B$^\mathrm{S/S}$ are shown in (a) with black solid line and 
 green dashed line, respectively. AB is shown in (b) with black solid line. 
 B$^\mathrm{Mo/S}$ and B$^\mathrm{Mo/Mo}$ are shown in (c) with black solid line and
 orange dashed line, respectively.
 In the same order and colors, (d), (e) and (f) show the band structures at their 
 equilibrium ILS.
 The blue or red shaded region marks the difference in VBM energy between 
  B$^\mathrm{Mo/S}$ and AA stackings.
}
\end{figure}

\begin{figure}
  \includegraphics[scale=0.84]{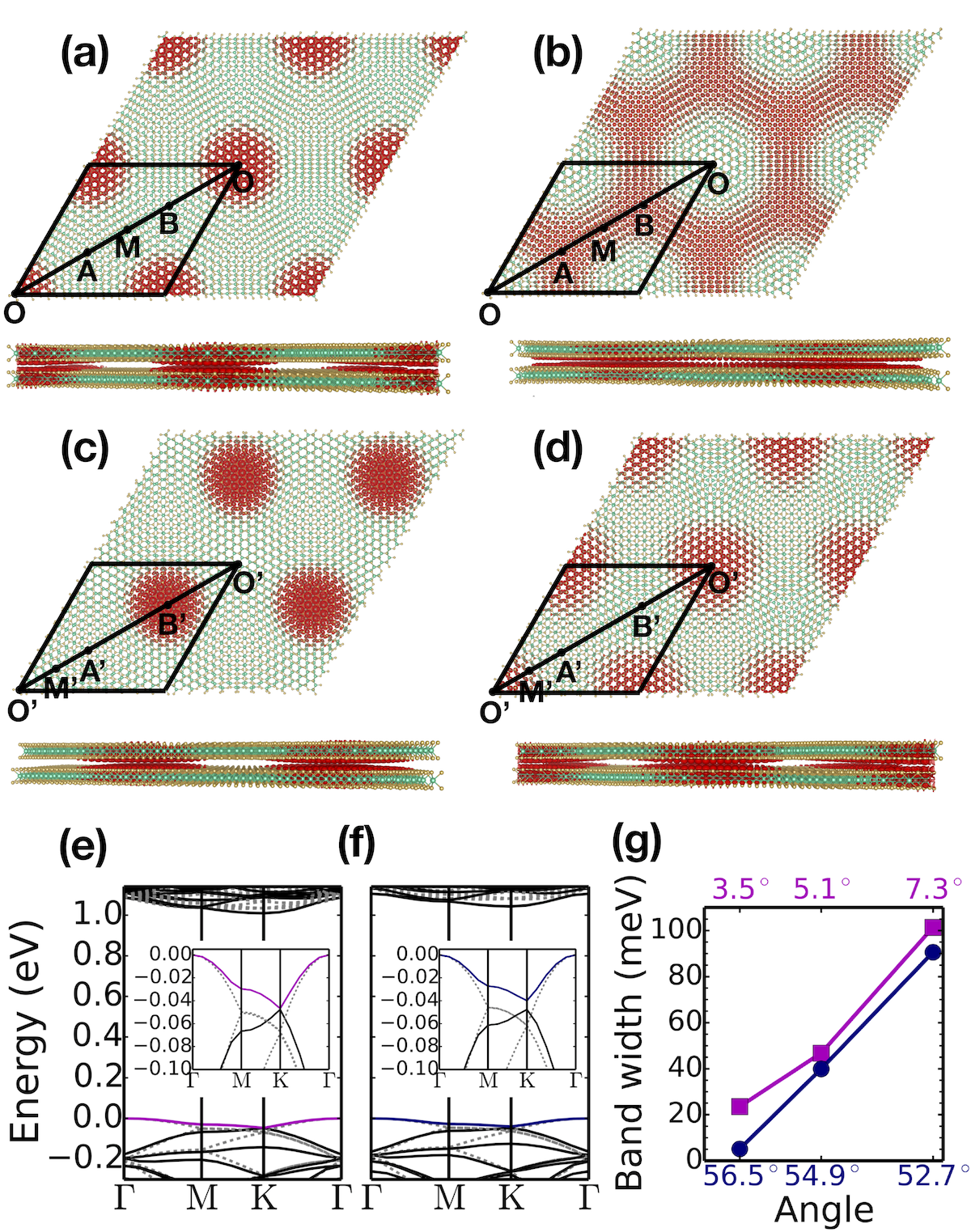}
 \caption{\label{fig4}
 (a) and (c) ((b) and (d)) Charge density of the VBM wavefunction in rigidly twisted (relaxed) 
  M$^{3.5}$ and M$^{56.5}$, respectively. 
  The isosurface value for these plots is $3\times10^{-4}$ $e/\mathrm{\AA}^{3}$.
 (e) and (f) Band structure of the relaxed M$^{5.1}$ and M$^{54.9}$, respectively. The flat bands,
 near the valence band edge are shown with magenta and blue colors, respectively. The dashed line 
 represents band structure of pure B$^\mathrm{Mo/S}$ and AB stacking for the same superlattice size, respectively.
 The inset shows an enlarged plot of the valence bands.
 (g) Variation of the band width (in meV) with twist angle. The magenta line corresponds to angles 
 approaching 0$^{\circ}$ and blue line to angles approaching 60$^{\circ}$.
}
\end{figure}

Among the band structures of stackings shown in Fig. \ref{fig3} (a), (b) and (c);
AA and B$^\mathrm{S/S}$ show the 
largest splittings at the $\Gamma$ point VBM (Fig. \ref{fig3} (a)).
This is due to the close proximity of the S atoms in these
stackings. 
Note that the ILS is fixed at 5.9 $\mathrm{\AA}$ for these band structures, 
same as that for rigidly tBLM.
The effect of these large splittings is that the VBM (with respect to the 
vacuum level) of the AA and B$^\mathrm{S/S}$ 
stackings is higher ($\sim$-4.8 eV) than the rest of the stackings ($\sim$-5.1 eV).
In the rigidly-twisted M$^{3.5}$, the unit cell band structure of each layer is folded into the 
Brillouin zone of the MSL (tBZ). 
Thus, the VBM of M$^{3.5}$, which contains both AA and B$^\mathrm{Mo/S}$ 
stackings, must have contribution from the local AA stacking region alone. 
This leads to the localization of the VBM wavefunction around the O point, as 
shown in Fig. \ref{fig4} (a). 
In a similar manner, 
 the VBM wavefunction of rigidly-twisted M$^{56.5}$ is localized to the B$^\mathrm{S/S}$ region or the 
B' point (Fig. \ref{fig4} (c)). 

The equilibrium ILS
in AA and B$^\mathrm{S/S}$ stackings is larger than other stackings (Fig. \ref{fig2} (a)).
This diminishes the hybridization between the two layers leading to a marked reduction in the 
splitting at the $\Gamma$ point VBM as shown in Fig. \ref{fig3} (d). 
These splittings 
are now smaller than 
the other stackings (Fig. \ref{fig3} (e) and (f)).
As a result, the VBM of the AA and B$^\mathrm{S/S}$
stackings is now lower 
than the rest of the stackings. 
The VBM in relaxed M$^{3.5}$ is then expected to originate from 
the local B$^\mathrm{S/Mo}$ and B$^\mathrm{Mo/S}$
regions. Fig. \ref{fig4} (b) shows the VBM wavefunction in M$^{3.5}$, 
which is indeed localized at these regions, forming a hexagonal network. 
The localization pattern is significantly different from the 
rigidly twisted case, demonstrating the important role of atomic relaxations.
The VBMs of 
AB and B$^\mathrm{Mo/Mo}$ stackings are close to each other (within $\sim$0.1 eV).
Based on the band structures, the wavefunction is expected to lie 
at the B$^\mathrm{Mo/Mo}$ region in the MSL, since this stacking 
has a higher VBM level than AB. But the opposite localization is 
found in M$^{56.5}$, shown in Fig. \ref{fig4} (d), where the VBM is restricted to the 
AB (O' point) stacked regions. 
This is due to small tensile and compressive strains 
in the local B$^\mathrm{Mo/Mo}$ and 
AB regions, respectively. On taking this strain ($\sim$0.3$\%$) into account, the order of 
the VBM with respect to the vacuum level is reversed (see SM). 
The VBM thus localizes completely to the AB stacked regions (Fig. \ref{fig4} (d)).
Furthermore, the CBM with respect to the vacuum level lines up among the stackings in 
Fig. \ref{fig3} (d), (e) and (f). Hence,
no localization is found close to the CBM for M$^{3.5}$ and M$^{56.5}$

The localization of the VBM in the MSL is accompanied by flattening of the band in the tBZ. 
The band structure for tBLM in the tBZ for twist-angle 5.1$^\circ$, M$^{5.1}$, and 54.9$^\circ$, 
M$^{54.9}$, is shown in Fig. \ref{fig4} (e) and (f). The Moir\'e bands close to the valence band edge are 
flatter than their pristine counterparts. Furthermore, a band gap opens at the $K$ point in the tBZ for twist 
angles close to 60$^\circ$. We define band width, W, for the top valence band between the $\Gamma$ 
and $K$ point (see SM for comparison of the hole effective mass). 
Fig. \ref{fig4} (h) shows the trend in W with twist angle. An ultraflat 
band with W = 5 meV is formed in M$^{56.5}$, separated in energy from other valence
states by 60 meV. 
The W for M$^{3.5}$ is larger due to larger extent of real space localization, 
ie. the formation of hexagonal networks rather than a spot. 


We also estimate the on-site Coulomb interaction, U, for M$^{56.5}$ to be $\sim$220 meV.
Computed using $U = e^2/(4 \pi \epsilon d)$, where d = 22 $\mathrm{\AA}$ from the
charge density of the localized state and
in-plane dielectric constant, $\epsilon$ = 3 (see SM).
A large ratio for U/W suggests the possibility of a Mott insulator phase at half-filling of the band \cite{Nature.Cao2}.
As discussed above, the 
conduction bands show no localization close to the Fermi level. The CBM is two-fold degenerate 
and delocalized in the 
MSL with weak interlayer hybridization.
Hence, an external electric field in the out-of-plane direction can easily split these bands 
(see SM) and localize the CBM onto one of the layers \cite{NL.Ting}. 

\begin{table}
\centering
\begin{tabular}{c@{\hskip 0.14in}r@{\hskip 0.08in}r@{\hskip 0.08in}r@{\hskip 0.08in}r@{\hskip 0.08in}r}
\hline
\hline
MX$_2$ & AA & B$^\mathrm{X/X}$ & AB & B$^\mathrm{M/X}$ & B$^\mathrm{M/M}$
         \\
\hline
MoSe$_2$ & -5.06 & -5.07 & -4.70 & -4.68 & -4.73 \\
WS$_2$   & -5.34 & -5.34 & -4.93 & -4.90 & -4.87  \\
WSe$_2$   & -5.02  & -5.04  & -4.67  & -4.73 & -4.62   \\
\hline
\hline
\end{tabular}
\caption{ 
VBM (in eV), with respect to the vacuum level, for the five stackings in 
other transition metal dichalcogenides: MoSe$_2$, WS$_2$ and WSe$_2$. M stands for the 
transition metal and X for the chalcogen.
}
\end{table}

We find that the band width, localization of the flat band and atomic relaxations do not change if
a different exchange-correlation functional is used in the DFT calculations (see SM). 
We also find that the relative ordering of the VBM among the stackings  which determines the localization
remains the same in GW calculations. \cite{PRB.Hybertsen,hedin65} (see SM)
Furthermore, we show that this feature is generic to other TMDs (MX$_2$)
by computing 
the VBM with respect to the vacuum level for the five high-symmetry stackings at the DFT level. 
The results are 
shown in Table I (see SM for band structures) for the equilibrium ILS.
For all TMDs, the AA and B$^\mathrm{X/X}$ stackings have VBM levels about $\sim$ 0.3 eV below the VBMs of 
other stackings. The CBMs, on the other hand, line up among the stackings. 
We do not expect spin-orbit coupling to significantly alter our conclusions (see SM).
Hence we posit that a similar localization should occur at the valence band edge of these TMDs, in 
close resemblance to what we have shown for MoS$_2$.  

In conclusion, we show the formation of ultraflat electronic bands close to the valence band
edge in MSLs of tBLM. Our analysis of the origin of the flat band indicates that twisted
bilayers of other TMDs must also show a flat band at the valence band edge.
The spatial-localization of electrons at the valence band edge will influence the binding energy
and dynamics of excitons. The spatially varying band gap could lead to the formation of exciton
funnels \cite{NL.Wu}.
Doping the flat band with holes could lead to spin-liquid states, quantum anomalous Hall insulators,
Mott-insulating phases, etc. at special filling factors \cite{arxiv.Mac}. 
Furthermore, the localization pattern of the flat band can be tuned with twist angle,
and is determined by atomic relaxations in the Moir\'e pattern.
The solitons can be probed through scanning tunnelling microscopy
and transport measurements, and could host topological edge states at small twist angles \cite{NC.Yin,PRX.Vaezi,Nature.Ju}.

\begin{acknowledgments}
The authors thank Sumilan Banerjee, Arindam Ghosh and H. R. Krishnamurthy for useful discussions. 
The authors thank the Supercomputer Education and Research Centre (SERC) at IISc
for providing the computational facilities.
\end{acknowledgments}



%
\end{document}